\documentclass[preprint,authoryear, 3p]{elsarticle}
\usepackage{pgfplotstable}
\usepackage{booktabs}
\usepackage{longtable}
\usepackage{array}
\usepackage{url}
\usepackage{amssymb}
\usepackage{natbib}
\newcolumntype{R}[1]{>{\raggedleft\arraybackslash}p{#1}}

\begin{document}

\begin{frontmatter}

\title{A Survey of Low-Velocity Collisional Features in Saturn's F Ring}
\author[att]{Nicholas O. Attree\corref{cor1}}
\ead{N.O.Attree@qmul.ac.uk}
\author[murr]{Carl D. Murray}
\author[will]{Gareth A. Williams}
\author[coo]{Nicholas J. Cooper}

\cortext[cor1]{Corresponding author}

\address{School of Physics and Astronomy, Queen Mary University of London, Mile End Road, London E1 4NS, UK}

\begin{abstract}
Small ($\sim50$km scale), irregular features seen in Cassini images to be emanating from Saturn's F ring have been termed mini-jets by \citet{attree12}. One particular mini-jet was tracked over half an orbital period, revealing its evolution with time and suggesting a collision with a local moonlet as its origin. In addition to these data we present here a much more detailed analysis of the full catalogue of over 800 F ring mini-jets, examining their distribution, morphology and lifetimes in order to place constraints on the underlying moonlet population. We find mini-jets randomly located in longitude around the ring, with little correlation to the moon Prometheus, and randomly distributed in time, over the full Cassini tour to date. They have a tendency to cluster together, forming complicated `multiple' structures, and have typical lifetimes of $\sim1$d. Repeated observations of some features show significant evolution, including the creation of new mini-jets, implying repeated collisions by the same object. This suggests a population of $\lesssim1$km radius objects with some internal strength and orbits spread over $\pm 100$km in semi-major axis relative to the F ring but with the majority within $20$km. These objects likely formed in the ring under, and were subsequently scattered onto differing orbits by, the perturbing action of Prometheus. This reinforces the idea of the F ring as a region with a complex balance between collisions, disruption and accretion.

\end{abstract}


\end{frontmatter}

\section{Introduction}
The F ring region has long been thought home to a population of small bodies. Evidence ranges from the depletion of magnetospheric charged particles measured by Pioneer 11 \citep{cuz88} to more recent stellar occultation results from Cassini VIMS and UVIS \citep{meinke12} as well as direct imaging by both Cassini ISS \citep{beurle} and the Hubble Space Telescope \citep{mcghee}.

At least some members of this population interact with the F ring core, with objects like S/2004 S 6 (hereafter referred to as S6) thought to produce large jets by physical collisions with the core \citep{cha05, mur08, cha09}. These jets extend over hundreds of kilometers in semi-major axis and persist for many days or months with new jets forming as S6 re-collides to form a sequence of features. These evolve through Keplerian shear, wrapping around through more than 360 degrees of longitude, to form the kinematic spiral strands seen on either side of the core \citep{cha05}.

The core also contains many small, irregular features and remains changeable on timescales ranging from hours to years \citep{porc05, french12}, suggesting additional processes because Keplerian shear should quickly smooth out any longitudinal variations. \citet{attree12} identified many of these `mini-jet' features, examining the time evolution of one, and proposing low velocity ($\sim1$ms$^{-1}$) collisions with local objects as the formation mechanism. Here we present a much more detailed study of the full population of mini-jets observed to date in an attempt to investigate the collisional theory and put constraints on the underlying moonlet population. 

Section 2 describes how the features are catalogued with measurements made from Cassini ISS images. In Section 3 we detail the basic orbital theory needed to explain the motion of mini-jets and outline the governing equations. Section 4 describes work done to analyse the distribution in space and time, proximity to Prometheus and evolution of the observed features. Some interesting examples are also discussed in more depth. Further analysis and the implications for the moonlet population are discussed in Section 5 before our concluding remarks are presented in Section 6.

\section{Observations}
A search of all Cassini ISS sequences containing resolved images of the F ring has been performed. As before \citep{attree12} we identified by eye small features (typically $\sim10$--$200$~km in radial extent and $\ll1^\circ$ in longitude) emanating from the core, excluding bright kinks and clumps within the core itself and the larger jet features. After pointing the images using background stars we re-project a portion of each image containing a feature in an equal aspect radius-longitude plot for direct comparison and then assigned them to one of three classes according to their morphology: (i) classic mini-jets, (ii) objects and (iii) complex features; each is described in detail below. Representative examples of each are shown in Figs.~\ref{fig1}, \ref{fig2} and \ref{fig3}. All co-ordinates are given in the standard Saturn centric reference frame with the $x$-axis corresponding to the position of the ascending node of Saturn's equatorial plane on the mean Earth equator at the J2000 epoch. Co-rotating longitudes are referenced to an epoch of 12:00 UTC, 1st January 2007 using the mean motion in section 3 below.

As of the end of 2012 we have catalogued 889 discrete features, including $32$ which are imaged more than once. Many are found clustered close together in `multiple' structures and we have tried to identify individual mini-jets within them, finding $56$ of these. In some cases, however, this proved impossible due to poor resolution or the number of individual features (sometimes dozens) and these were placed in the `complex' category below. Some examples of multiple features are shown in Fig.~\ref{fig4}.  Both repeated and multiple features are of special interest and are discussed in more detail, in Section 4.3, below.

The full catalogue is available online as supplementary material A. Entries are named by the Cassini image number that they appear in while multiple features in the same image are labelled alphabetically e.g. N1557026084a, N1557026084b etc.

\subsection{Classic Mini-Jets}

Features with a clearly defined linear structure at a measurable angle from the longitudinal direction are defined as classic mini-jets like the original feature discussed in \citet{attree12}. As shown in Fig.~\ref{fig1} they are seen at a variety of angles and lengths and a significant sub-category has been identified as `bright-heads'; these are classic mini-jets with a particularly bright tip or head. Such a head may represent the colliding object itself or simply a concentration of ejecta at that location. 437 mini-jets, including 37 bright-heads have been found.

\subsection{Objects}

`Objects' have been identified as bright features separate from, but close to, the F ring core. As seen in Fig.~\ref{fig2} most are extended longitudinally in a similar way to S6 and some are connected to the core by a faint dust sheet or, in some cases, a mini-jet-like linear trail. The dividing line between objects with a faint trail and mini-jets with a bright head is therefore somewhat blurred and it is possible that these actually represent the same phenomenon imaged at different geometries or orbital phases. 207 objects have been identified.

\subsection{Complex Features}

As shown in Fig.~\ref{fig3} some features are difficult to fit into either of the above classes and are instead described as `complex'. Some are most likely mini-jets or objects imaged at poor resolution or difficult geometries but we suggest that many are more complex features. A number of these appear in the very highest resolution images, for example there are 14 in one early sequence (ISS\_029RI\_AZSCNLOPH001\_PRIME) with just over $1$km radial resolution and $10^{-3}$ degrees longitudinal resolution. These may reveal a complicated substructure present in all mini-jets but not normally resolved. Several others have the appearance of a superposition of several classic mini-jets on top of one-another and resemble close multiple features. These are placed in the complex category when the resolution is insufficient to separate the individual components. Altogether we have found 245 complex mini-jets.

\section{Theory}
We define the variables $\Delta a$ and $\Delta e$ as the differences in semi-major axis and eccentricity, respectively, from an orbit with a certain $a$ and $e$. Mini-jets similar to the original feature consist of material with a range of $\Delta a$ and $\Delta e$, with $\Delta a \approx a\Delta e$, ranging from 0 at the base up to some maximum at the tip \citep{attree12}. Epicyclic theory (see e.g. \citealp{Murray}) predicts that, in a frame co-rotating with our start point (with $a$,$e$), particles in the mini-jet will each follow their own $2:1$ centred ellipse with the addition of a linear drift due to Keplerian shear. All the particles move in phase to create the coherent linear structure. Radial, $r$, and longitudinal, $l$, positions in kilometers relative to the base will, to first order, follow
\begin{equation}
r = \Delta a - a \Delta e \cos M,
\qquad
l = 2a\Delta e \sin M - 3/2M\Delta a
\label{r,l}
\end{equation}
where we take $a=140221.3$km as the semi-major axis of the F ring core \citep{albers12}, $M=nt$ is orbital phase in radians, with $n=10.157$ radians per day the mean motion, and $t$ is time. The particles start off at periapse for mini-jets with positive $\Delta a$ and aposapse for those with negative $\Delta a$.

As shown in a simple schematic (Fig.~\ref{fig5}) the tangent of the angle that the mini-jet makes with the horizontal is simply the ratio of these two lengths, i.e.
\begin{equation}
\theta = \tan^{-1}\frac{1 - \cos M}{2 \sin M - 3/2 M}
\label{theta}
\end{equation}
where, as noted above, we use the relationship $\Delta a \approx a\Delta e$. Equation (\ref{theta}) is plotted in Fig.~\ref{fig6} and describes the time evolution of the angle of any mini-jet, irrespective of size, as long as the above relation holds. Note that this is the same as the cant angle from \citet{tisc13}. Also shown in Fig.~\ref{fig6} are the measured angles for the original feature \citep{attree12} plotted against time, with the start time shifted to best fit the curve. The image numbers and measured co-ordinates of the mini-jet tip and base (from which the angles were calculated) can be found in supplementary information B.

The scale of the mini-jet is determined by the speed of the collision which formed it; the maximum $\Delta a$ being therefore related, by the standard perturbation equations, to the velocity `kick', $\Delta \bf v$,  delivered to the particles at the tip. Under certain assumptions (dissipative collisions between F ring particles and a much more massive impactor, see \citet{attree12}) this is of the same order as the collision velocity, $\bf U$, between the impacting object and the target (the F ring core or strand). The component of $\Delta \bf v$ in, or opposite to, the direction of orbital motion dominates over any radial components so we can simplify the perturbation equations to
\begin{equation}
U \sim \Delta v \approx \frac{n\Delta a}{2}
\label{U}
\end{equation}
In essence the tip of the mini-jet is put onto an orbit approximating that of the colliding object while the rest of the material is dragged out between this and the F ring.

It seems likely that the complex features are also collisional in nature but with more complicated ejecta patterns due to multiple objects or differing collision geometries. We have yet to develop a general theory for their time evolution and instead analyse them on an ad-hoc basis.

\section{Analysis}

For classic features we measured the radial and longitudinal co-ordinates of the tip and base of the mini-jet and computed the lengths $r$ and $l$ and the angle $\theta$ from these. For complex features and objects we measured the co-ordinates of the center of the feature (labeled base co-ordinates in supplement A) and the center of the object (labeled tip co-ordinates), respectively. Uncertainties in these measurements vary with image geometry and resolution, with a typical pixel being $5$km in the radial and $0.01^\circ$ in the longitudinal direction. This pixel error is carried forward when calculating the angle $\theta$ but the blurred, extended appearance of most features means it is an underestimate and we instead estimate typical errors by eye to be $\sim \pm2^\circ$. In the analysis below the derived quantities should then be taken as order of magnitude estimates.

\subsection{Distribution}

A total of 857 unique features (see repeats in Section 4.3 below) have been found in 110 image sequences. The number in each sequence is highly variable, ranging from zero to $47$, with a median of $5$ and a mean and standard deviation of $8 \pm 8.6$ respectively. Features are found on both sides of the core and in both inner and outer strands ($\sim60$ in the strands) and at a range of phase angles ($13^\circ-163^\circ$) and viewing geometries.

Examining the numbers of features at each longitude in a frame co-rotating with the F ring revealed no noticeable trends (in an inertial frame the number is biased to where Cassini makes its observations). Likewise, when examining the Fourier transform of the distribution no traces or periodicity were detected. No particular trends were detected within the subclasses of feature and no particular differences between their distributions have been noted.

Mini-jets and similar features, then, are distributed randomly around the F ring, as might be expected from a stochastic, collisional process. The average separation between adjacent features in the same sequence is $22^\circ \pm 23.0^\circ$, where the error is the standard deviation which is large because the number of features in each sequence is so variable. We tested for clustering by taking the ratio of this to the expected `degrees per feature' (the sequence coverage divided by number of features seen in it) and found that for nearly all sequences this was around one (median for the whole data set of $1.17$). A value of one would be expected for no clustering as the observed average distance would match that of uniformly spaced features (e.g. following a Poisson distribution with a mean separation). On the large scale at least, the distribution is fairly uniform around the ring. On the small scale, however, there are over 50 `multiple' features where the separation between adjacent features is $\ll1^\circ$, much less than the average. This suggests that the components of `multiples' are related to each other rather than being randomly distributed mini-jets that happened to be seen close together. For features with a mean separation of $\sim 20^\circ$ the probability of finding two $<1^\circ$ apart is given by Poisson statistics as negligible ($\sim10^{-9}$) even with over $800$ features.

Finally, in terms of spatial distribution, the longitude of each feature relative to Prometheus is plotted in Figure \ref{fig7}. The mean number of features in each $2^\circ$ bin is $5 \pm 2.66$. The distribution is generally random and noisy but, with half of the bins within $10$ degrees of Prometheus more than one standard deviation less than the mean, there is a slight dip in the number of features here.  One might expect mini-jets to be more difficult to observe in the highly disturbed streamer-channel region and this may account for the dip, however there is some evidence that the decrease persists upstream of Prometheus, where the ring should be relatively undisturbed. In this case the feature would be real although the physical mechanism for a depletion in collisional features near Prometheus is, as yet, unexplained (one possibility is a phase-lag effect as in the `predator-prey' model of \citealp{esp12}). When considering the scatter in the rest of the data set we suggest that the decrease is around the level of the noise and any trend here is weak at best.

The time distribution of the features was then examined by plotting the total number in each observation sequence, adjusted by longitude coverage, against date. This is shown in Fig.~\ref{fig8}. Adjusting for the varying amount of the F ring covered in each sequence is done by dividing the number of features by the co-rotating longitude coverage and then multiplying by 360 degrees to get the expected number in the whole F ring at any one time. This is reasonable if features are distributed randomly so that all longitudes should contain a similar numbers of mini-jets. Only sequences with a co-rotating longitude coverage of $> 50^{\circ}$ are deemed representative of the whole ring and are included in this plot. A similarly shaped graph to Fig.~\ref{fig8}, with the same trends, is seen when plotting just the raw number of features over time suggesting the validity of the adjustment method. The mean and standard deviation of the adjusted number of features in these sequences is $14 \pm 13.1$ respectively. This agrees with the median, adjusted number for the whole data set of $16$ and is just within the limits of the mean of the raw count, quoted above, which is weighted downwards by the large number of small longitude coverage observations with a single feature. In addition $16$ features in the whole F ring would match the average of $22^\circ$ per feature mentioned above. Thus we take $\sim 15$ as our average for the number of features visible in the F ring at any one time.

The symbols in Fig.~\ref{fig8} are coded by the range from the spacecraft to the ring as a measure of resolution, we find only a weak trend for more features at better resolution. Indeed the number of features is highly variable, as seen in the large standard deviation, and not particularly well correlated with any other observing statistic. This is partly because mini-jets are only visible at certain points in their orbit, depending on phase and geometry, and partly because of the difficulties of visual identification, especially when confronted with multiple features.

The date of closest approach between the F ring and Prometheus (due to their eccentricities and differing precession rates) is also highlighted in Fig.~\ref{fig8} and there is a hint of an increase in the number of features in the year leading up to it. Unfortunately there is little coverage of the F ring at the actual closest approach and after it and any increase preceding it must be treated with caution because of the large number of observations here. The 2008/09 observations are among the best in terms of coverage, number and resolution and while we have tried to remove any biases by the methods above it should, perhaps, come as no surprise that the best observations produced the most entries in the catalogue. As above, when considering the large scatter, we suggest that the increase in 2008/09 is at the level of the noise ($6$ sequences above $2$ standard deviations and none above $3$) and any trend for greater numbers of mini-jets here is weak at best.

\subsection{Morphology}

For the subset of 437 `classic' and 'bright-head' mini-jets we measured the lengths $r$ and $l$ and the angle $\theta$. We then used a graphical method to find all the points where the curve from  Eqn.~(\ref{theta}) intersected this measured angle, each one representing a possible solution. Equations (\ref{r,l}) and (\ref{U}) were then used to find the $\Delta a$ and an approximate collision velocity for each solution.

As can be seen in Fig.~\ref{fig6} for small angles ($< 8.58^{\circ}$) there can be many such possible solutions as the mini-jet could be collapsing on its first cycle or rising on its second or subsequent cycle. There is some degeneracy here e.g.~a large, young mini-jet can appear similar to a small, old one. We computed an age and $\Delta a$ for each solution and then used the second part of Eqn.~(\ref{r,l}) to predict a length $l$ for each. These predicted lengths were then compared to the measured length and the best fit was chosen as the preferred age/$\Delta a$ solution. In effect this is a best guess at the real properties of the mini-jet and is the best that can be done from a single image. We checked for self-consistency by comparing the predicted lengths to the maximum and minimum lengths that a mini-jet of that $\Delta a$ and number of cycles could be and found all chosen solutions to fit. Nonetheless we still found a small number ($18$) of derived $\Delta a$ values to be very large and consider these outliers to be poorly fitted by the above method. These were all young, `wrong-way' (against the direction of Keplerian shear) or large, jet-like, features which are difficult to measure but the fact that they are all much too large (hundreds of km) may mean we have misunderstood the early phases of mini-jet formation.

Figure \ref{fig9} shows the distribution of measured angles as well as a typical predicted distribution for comparison. This was created by generating a population of mini-jets with random ages (weighted by number of cycles as found from the best fit ages) then using Eqn.~(\ref{theta}) to produce a predicted angle for each. This serves to highlight the observational bias inherent in detecting mini-jets by eye: we find many more at medium angles ($10^{\circ}-50^{\circ}$) and many fewer at low angles ($<10^{\circ}$) than predicted, simply because low angles are harder to spot in amongst the irregular F ring core. Apart from this discrepancy the measured angles are consistent with a population of mini-jets with ages ranging from a few hours to a few days although fewer young, `wrong-way' mini-jets are detected than might be predicted. This is probably because they are very difficult to measure but again see the above comment on outliers.

Figure \ref{fig10} shows the best fit solutions for $\Delta a$ omitting these outliers. Nearly all the mini-jets derived with confidence lie within $\Delta a = 100$km of the core with roughly equal numbers positive and negative and a mean absolute $\Delta a = 24 \pm 17.2$km (error is one standard deviation). This is a collision speed of $U \sim1.2$ms$^{-1}$. The corresponding ages have a mean of $1.27$ orbital periods or $0.79$ days and a little over a quarter (125/437) of mini-jets appear to have survived more than one cycle suggesting $\sim3/4$ are disrupted when re-entering the core at the end of their first cycle.

\subsection{Multiple and Repeated Features}

The multiple mini-jets in Fig.~\ref{fig4}a and c have similar angles, $\theta$, and similar sizes implying that  they formed at roughly the same time with the same collision velocities. A cluster of objects on very similar orbits, but stretched out in longitude by Keplerian shear, would impact the ring in a chain creating just such a series of mini-jets. We consider these to be evidence for groups of objects or objects disrupted into swarms.

By contrast Fig.~\ref{fig4}d to g have multiple mini-jets of different angles (ages) where the tips of each jet all seem to point to the same spot (Fig.~\ref{fig4}d has three features showing both types). This would be expected if an object survives to collide with the core again at the next loop (see Fig.~\ref{fig5}) creating a second mini-jet, in phase with, and following the tip of the first. Thus the objects in Fig.~\ref{fig4}d to g may be re-colliding one orbital period later. Below we discuss repeated observations of the same features which further supports this idea,

Nearly all of our catalogued features are seen in only $2$ to $3$ images as the F ring rotates through the ISS field of view over the course of a few minutes. However some observation sequences are taken within hours or days of each other and cover the same section of ring (i.e.~the same range of co-rotating longitude) raising the possibility of repeated observations of the same feature. In principle this means more data points to fit to the curve in Fig.~\ref{fig6}, allowing better constraints on mini-jet properties, whilst also directly measuring their lifetimes and any subsequent collisions.

To this end we searched the catalogue by co-rotating longitude and date before comparing the morphology of features which are close in space and time. We have found 25 possible matches,  imaged between $\sim15$ hours and $\sim6.7$ days apart, including 63 individual features which are listed in Table \ref{tab1}. There is no special distribution in time or longitude in those features which are repeated suggesting that it is only due to chance and good coverage that we managed to observe them.

Visually the repeated features appear more sheared, as expected, being both longer (larger $l$) and at a lower angle (smaller $\theta$), however they proved difficult to fit to the predicted evolution of $\theta$. As noted in the table many are deemed complex with large uncertainties in their angles and morphologies and no particularly good fits were achieved, while classic mini-jets imaged more than one cycle apart also proved difficult to fit. This might imply that morphological changes occur when a mini-jet re-enters the core, e.g.~the release of more ejecta or re-collision and displacement onto differing orbits. Another possibility is that the morphologically similar feature seen at a later date actually represents a new collision, at a similar geometry, after the first feature has dissipated. The extent to which colliding objects survive and go on to re-collide provides information about their physical properties and those of the core.

\small{
\begin{longtable}{p{1.1in} p{1.5in} R{0.2in} p{1.2in} R{0.5in} R{0.4in}}
\centering Designation & \centering Date (year-DOY-time)& \centering $\theta$($^{\circ}$) & \centering Class & \centering Phase Angle($^{\circ}$) & \centering Exposure (ms) \endhead
\hline
N1538204682 & 2006-272-06:32:50.814 & -- & Extended Object & 161 & 680\\
N1538283887 & 2006-273-04:32:55.309 & -- & Extended Object & 160 & 680\\
\hline
N1557026084a & 2007-125-02:40:53.219 & $-46$ & Classic - bright head & 80 & 1000\\
N1557026084b & 2007-125-02:40:53.219 & $-18$ & Classic - bright head & 80 & 1000\\
N1557080024 & 2007-125-17:39:52.877 & $-11$ & Classic - bright head & 83 & 1000\\
\hline
N1577828765a & 2007-365-21:09:57.025 & $-1$ & Classic & 65 & 1200\\
N1577828765c & 2007-365-21:09:57.025 & $-4$ & Classic - bright head & 65 & 1200\\
N1578409567a & 2008-007-14:29:55.038 & $-2$ & Classic - bright head & 21 & 1000\\
N1578409567b & 2008-007-14:29:55.038 & $0$ & Classic & 21 & 1000\\
\hline
N1595326037 & 2008-203-09:29:05.444 & -- & Extended Object & 104 & 1500\\
N1595501977 & 2008-205-10:21:24.905 & $-24$ & Classic & 13 & 100\\
\hline
N1606006222 & 2008-327-00:10:55.278 & $-55$ & Classic & 38 & 1200\\
N1606033182 & 2008-327-07:40:15.087 & $-43$ & Classic - bright head & 46 & 1200\\
\hline
N1623224391a & 2009-160-06:58:20.565 & $-43$ & Classic & 95 & 1800\\
N1623224391b & 2009-160-06:58:20.565 & $-32$ & Classic & 95 & 1800\\
N1623332218a & 2009-161-12:55:27.199 & $-8$ & Classic & 17 & 1000\\
N1623332218b & 2009-161-12:55:27.199 & $-8$ & Classic & 17 & 1000\\
\hline
N1623224706a & 2009-160-07:03:35.563 & $-17$ & Classic & 95 & 1800\\
N1623224706b & 2009-160-07:03:35.563 & $-4$ & Classic & 95 & 1800\\
N1623332896a & 2009-161-13:06:45.194 & $-13$ & Classic & 17 & 1000\\
N1623332896b & 2009-161-13:06:45.194 & $-3$ & Classic & 17 & 1000\\
\hline
N1623225651a & 2009-160-07:19:20.556 & -- & Complex & 94 & 1800\\
N1623333348a & 2009-161-13:14:17.190 & -- & Complex & 17 & 1000\\
\hline
N1623225651b & 2009-160-07:19:20.556 & --& Complex & 94 & 1800\\
N1623360620 & 2009-161-20:48:48.597 & $26$ & Classic & 34 & 1800\\
\hline
N1623226701 & 2009-160-07:36:50.548 & -- & Complex & 93 & 1800\\
N1623335156b & 2009-161-13:44:25.178 & -- & Complex & 17 & 1800\\
\hline
N1623329280d & 2009-161-12:06:29.219 & -- & Extended Object & 17 & 1000\\
N1623355648 & 2009-161-19:25:56.632 & -- & Extended Object & 34 & 1800\\
\hline
N1623330636b & 2009-161-12:29:05.210 & $-40$ & Classic & 17 & 1800\\
N1623357230a & 2009-161-19:52:18.621 & -- & Complex & 34 & 1800\\
\hline
N1623330636c & 2009-161-12:29:05.210 & -- & Complex & 17 & 1000\\
N1623357230b & 2009-161-19:52:18.621 & -- & Complex & 34 & 1000\\
\hline
N1623342840 & 2009-161-15:52:29.123 & -- & Extended Object & 19 & 1000\\
N1623369434 & 2009-161-23:15:42.534 & -- & Complex & 34 & 1800\\
\hline
N1623345778b & 2009-161-16:41:27.102 & -- & Extended Object & 19 & 1800\\
N1623372372 & 2009-162-00:04:40.513 & $-23$ & Classic - bright head & 35 & 1800\\
\hline
N1629353989 & 2009-231-05:37:35.278 & $-13$ & Classic & 119 & 1200\\
N1629517865 & 2009-233-03:08:50.432 & $-5$ & Classic & 126 & 560\\
\hline
N1729028870 & 2012-289-20:54:33.308 & -- & Extended Object & 138 & 1800\\
N1729057850 & 2012-290-04:57:33.424 & -- & Extended Object & 155 & 1200\\
\hline
N1729030994 & 2012-289-21:29:57.295 & -- & Complex & 138 & 1800\\
N1729059974a & 2012-290-05:32:57.411 & -- & Complex & 155 & 1200\\
N1729059974b & 2012-290-05:32:57.411 & -- & Complex & 155 & 1200\\
\hline
N1729035773 & 2012-289-22:49:36.264 & $-1$ & Classic & 139 & 1800\\
N1729064753 & 2012-290-06:52:36.380 & $-2$ & Classic & 156 & 1200\\
\hline
N1729037366a & 2012-289-23:16:09.254 & $-30$ & Classic & 139 & 1800\\
N1729037366b & 2012-289-23:16:09.254 & $-17$ & Classic & 139 & 1800\\
N1729259467a & 2012-292-12:57:49.144 & $-21$ & Classic & 71 & 1200\\
N1729259467b & 2012-292-12:57:49.144 & $-6$ & Classic & 71 & 1200\\
N1729259467c & 2012-292-12:57:49.144 & $-9$ & Classic & 71 & 1200\\
\hline
N1729043207 & 2012-290-00:53:30.217 & -- & Complex & 140 & 1800\\
N1729072187 & 2012-290-08:56:30.333 & $-14$ & Classic & 158 & 1200\\
\hline
N1731113813 & 2012-314-00:03:23.367 & -- & Complex & 152 & 1200\\
N1731140093 & 2012-314-07:21:23.200 & -- & Complex & 145 & 1200\\
\hline
N1731126143 & 2012-314-03:28:53.288 & $-1$ & Classic & 155 & 1200\\
N1731152834 & 2012-314-10:53:44.119 & $0$ & Classic & 148 & 1200\\
\hline
N1731129842a & 2012-314-04:30:32.265 & -- & Complex & 156 & 1200\\
N1731129842b & 2012-314-04:30:32.265 & $-14$ & Classic & 156 & 1200\\
N1731157355 & 2012-314-12:09:05.090 & -- & Complex & 149 & 1200\\
\hline
N1734573588 & 2012-354-01:05:56.394 & $-54$ & Classic & 96 & 1200\\
N1734595367a & 2012-354-07:08:55.255 & $-12$ & Classic & 91 & 1200\\
N1734595367b & 2012-354-07:08:55.255 & $-3$ & Classic & 91 & 1200\\
\caption{Repeated features, grouped together (each set separated by horizontal lines) and ordered by time of first appearance. Some new features have appeared on the second observation and these are given new designations, whilst others have disappeared or merged. The time between observations ranges from $\sim15$ hours to $\sim6.7$ days and is noted in Day Of Year format in the second column. Classes are determined by appearance as described in the observation section and the observation phase angle and camera exposure time are also noted. The angle $\theta$ is measured from the tip and base co-ordinates as described above and errors are generally $\sim 2^{\circ}$.}\\
\label{tab1}
\end{longtable}}

\normalsize{
We now examine one particular repeated feature, seen in May 2007, which presents an interesting challenge for mini-jet theory. Seen at first (top left of Figure \ref{fig11}) as a large double mini-jet with a bright head, it has evolved to a triple mini-jet by the time of the second image, approximately one orbital period later (bottom left of Fig.~\ref{fig11}). Also visible in Fig.~\ref{fig11} are a number of dark, sheared channels in the strands either side of the core up- and down-stream of the mini-jets, two in the first image and four in the second. We believe all these features can be explained by multiple collisions with a single object if, on colliding with the dense, clumpy core, it creates a mini-jet but while moving through the more diffuse strands it merely sweeps up or scatters material leaving behind a dark channel. We note that this kind of behavior can also occur with the larger jets and S6 \citep{mur13}.

The right-hand panels of Fig.~\ref{fig11} show corresponding frames from an animation of a simple model where mini-jets and channels are generated artificially to match the angles and locations of the observed features. Mini-jets are created by assigning a $\Delta a$ and $a\Delta e$ to particles with a random number generator, up to a maximum of the object's orbit (below), which then evolve according to Eqns.~(\ref{r,l}) and (\ref{theta}). Channels are likewise generated, but without any $a\Delta e$, as particles which are coloured white. A single object will pass through these locations in the right sequence at roughly the correct time to form these features if it has $\Delta a \approx 25$km, $a\Delta e \approx 110$km and $\Delta i \approx 10^{-3}$ degrees, where $\Delta i$ is the inclination relative to the core. For the object to pass through strands both inside and outside of the core it must have a relative eccentricity greater than its $\Delta a$, i.e. $\Delta a \ne a\Delta e$ contrary to what has been assumed for all mini-jets above. As far as we can tell this is not ruled out by the dynamics but could have implications for the derived mini-jet ages if some of them deviate from Eqn.~(\ref{theta}). Indeed it may help explain the poorly fitted outliers mentioned above if these are, in fact, eccentric mini-jets on the `wrong' side of the ring. This relative eccentricity also means that the object will pass through the core twice per cycle and a relative inclination is needed to explain why only one mini-jet is formed: the object loops above or below (i.e.~out of the orbital plane) the core in one direction and collides with it in the other. The three dimensional nature has not been considered until now because inclinations are very small ($i = 0.0067^{\circ}$) in the F ring but collisions and gravitational interactions should induce small inclination differences between objects. To a first approximation one might expect Prometheus to induce inclinations comparable to its own of $i = 0.008^{\circ}$. An inclination difference of only $\Delta i = 8.17 \times 10^{-4}$ degrees is needed to produce a vertical offset of (converting from radians) $z \approx 180a\Delta i/\pi = 2$km at the F ring, enough for $\lesssim1$km radius objects to `miss' each-other and pass by without colliding. Even such an initially complex feature as this can then be accounted for by a single colliding object with a few reasonable assumptions.

We present one final example of a mini-jet that has been imaged twice, one orbital period apart, in early 2013. This feature has not yet been included in the catalogue but is discussed here for comparison with the one above. The re-projected images are shown in Fig.~\ref{fig12} and the tip of the feature can be seen to have moved about $600$km between frames. This, and the angles, are consistent with a $\Delta a \approx 65$km mini-jet, the theoretical path of which is superimposed on the figure. The tip and/or colliding object itself must have re-entered the F ring core in the time between images and one might expect a new mini-jet, in phase with the original, to have formed as above. Instead, visible in the second image is a dark, sheared channel in the core located roughly at the point of re-collision. We suggest that this is a re-collision feature, as above, but one which happened to occur in an under-dense section of ring thus resulting in the sweep-up and scattering of fine grained material, leaving a depleted dark channel, rather than collision with a solid clump leaving a mini-jet. Finally we note that the appearance of the F ring before collision is uniform and bright so it is not immediately apparent that this is an under-dense region. This shows the difficulty in assessing the `clumpiness' of the core from images.

\section{Discussion}

The collisional theory proposed by \citet{attree12} remains consistent with the presented analysis and suggests a population of small bodies on orbits similar to the F ring's ($\sim10$s of km difference in semi-major axis) interacting with it. These would have to be $\sim1$km or less in diameter in order not to be resolved in ISS imaging although the objects listed in the observations section may represent some of the larger ones or those with accompanying dust envelopes.

Evidence exists for repeated collisions, suggesting that at least some of these objects survive passing through the F ring and so they must have some degree of strength, i.e.~they are not just clumps of material on similar orbits. Repeated collisions do seem rare but whether this is due to weak objects or an unlikely combination of orbits in a dense region of the highly variable F ring is difficult to say. Multiple impact features are also relatively common, suggesting groups of objects moving together on related orbits. Similar complex structure is visible in the larger S6 collisions and is also highly suggestive of clusters of objects. These likely represent those objects that have been partially disrupted into a swarm of smaller objects, either by previous collisions with the core or by tidal forces. Keplerian shear swiftly stretches out an initial clump into a long, extended chain.

\citet{tisc13} argue that elongated features observed in the A ring are dust clouds formed by impacts from cm- to m- sized meteoroids. They are fitted best as evolving purely by Keplerian sheer (rather than epicyclical motion combined with shear as in Eqn.~(\ref{theta})) and \citet{tisc13} suggest that this implies ``the impact of a compact stream'' of material from a broken up object in a similar way to our chains of mini-jets above. Though the dynamics are analogous we feel that the relative velocities and the presence of bright heads, objects and re-collision features presented in this paper suggest a local source for the F ring collisional population.

We can can crudely estimate the number of colliding objects with a particle in a box method using
\begin{equation}
N = \frac{2 \pi a H W \nu}{\sigma v},
\label{number}
\end{equation}
where $N$ is the required number of objects with average dispersion velocity $v \approx1$ms$^{-1}$ and collisional cross section $\sigma$ to result in a collision frequency of $\nu$ in a box of height $H$, width $W$ and length $2 \pi a $ (where $a$ is semi-major axis as before), taken to be $50 \times 50$km. The frequency of mini-jet forming collisions must be $\nu \sim 10$ day$^{-1}$ to sustain a visible population of that number, given their short lifespan of $\sim1$ day.

The collisional cross section between a continuous ring of width $x$ and a single spherical object of radius $r_{\rm{obj}} \sim 500$m is $\sigma = \pi r_{\rm{obj}}^2 + 2 \pi a x$, assuming collisions are equally likely from all directions, i.e.~taking the core as a flat `ribbon' of width $x$ all the way around its orbit. Taking $x = 10$km as the nominal core width results in $N \approx 30 $ objects. However the core is not uniform and simple, having dense clumps and variable width in longitude and time, not to mention a varying number of nearby strands. If we instead assume that significant features only form during collisions with a narrow, 1km, discontinuous inner core, containing the larger F ring particles \citep{mur08} then $N \approx 290$ objects are needed for the observed collision rate. If we further restrict mini-jet forming events to collisions between a finite number of discrete clumps or objects, rather than a continuous ring, we require more of these objects by several orders of magnitude; though the estimate will vary depending on how elongated the clumps are and how they move. Hence a more detailed knowledge of the collision process is needed to improve this estimate.

Previous attempts by \citet{cha09} to simulate the larger jet forming events have used two extremes: either a moonlet colliding with a continuous ring of massless particles or an unbound clump of massless particles colliding with an embedded moonlet. Of these the second is preferred as it produces jets of material on either side of the core centered on the moonlet as is often observed (see Fig.~6 of \citealp{cha09}). However it cannot be fully correct as jets are also seen on only one side of the core at a time and this is nearly always the case with mini-jets. Furthermore, an unbound group of particles would disperse and not go on re-colliding as S6 and some mini-jet objects clearly do. The true situation may be somewhere between these two extremes with loosely bound aggregates colliding with a complex, clumpy core.

For now, taking our number of features as a lower estimate because we may have missed low-angled mini-jets, we can say that the F ring region contains a steady population of at least $\sim100$ small objects interacting with the core. This would be a subset of the total number of objects implied by occultations which is estimated at several tens of thousands of objects greater than $100$m in size \citep{meinke12}.

Such a population would have to be continuously replenished as its members are eroded or broken down by collisions. This is further evidence for ongoing accretion in the F ring core as investigated by \citet{can95} and \citet{kar07} among others. There is already evidence that Prometheus may aid in triggering clump formation \citep{beurle, esp12} and perturbing objects onto their colliding orbits. Although collisional features are not correlated with Prometheus's location this does not rule out its influence as objects perturbed onto differing orbits would naturally spread around the ring.

Because the tip of a mini-jet is placed on an orbit similar to the colliding object, Fig.~\ref{fig10} approximates the distribution of the object population in semi-major axis relative to the core. For comparison the maximum $\Delta a$ perturbation exerted by Prometheus is $\sim20$km when it and the F ring are at closest approach \citep{mur08}. This means multiple encounters are necessary to place some objects onto their mini-jet forming orbits, if indeed this is the mechanism. Subsequent encounters would be equally likely to perturb an object back towards the ring, depending on its exact orbit and the phase of closest approach, meaning objects could undergo a chaotic random walk in $\Delta a$ around the core. Likewise their solidity and size would evolve by collisions in a random way over time with only a lucky few aggregating into solid moonlets as described in \citet{esp12}. We conclude this to be the likely formation mechanism for S6 which must be a relatively young object.

An overall picture then emerges of mini-jet forming objects, with some solidity and on orbits differing by tens of km from that of the core, being a subset of the total population of F ring objects. We consider these to be between temporary aggregates forming in the core and in the process of becoming larger more solid moonlets on more distant orbits like S6. Mini-jets and jets would then represent two ends of a continuum of collisional features formed by steadily increasing impact velocities from steadily increasing $\Delta a$ values. It appears that larger jets evolve purely under Keplerian shear rather than following Eqn.~(\ref{theta}) and would therefore have $\Delta a \ne a\Delta e$ as would their progenitor objects. Prometheus's perturbations should preserve $\Delta a \approx a\Delta e$ \citep{will09} so this might represent objects on more distant orbits freely precessing under Saturn's gravity rather than being locked to the F ring core. Further study of multiple Prometheus interactions and collisions on differing orbits will be needed to confirm this theory,

\section{Conclusion}
Small, irregular features have been found all around the F ring, throughout the time that Cassini has observed it. Their numbers are highly variable, but average $\sim15$ in the ring at any one time. Based on our analysis of nearly 900 catalogued features they are randomly distributed but with a tendency to clump together in multiple structures. There may be fewer near the location of Prometheus and more in the years leading up to closest approach with this satellite but these trends seem weak at best. Those with a resolvable linear structure (mini-jets) have angles consistent with a range of ages, from a few hours to a few days, and with around a quarter being less than one orbital period (15 hours) old. This is also supported by the repeated detection of several features over the course of a few days, suggesting an average lifetime of $\sim1$ day. Many repeated features also show significant morphological changes between images, including the creation of new mini-jets, emphasising their extremely dynamic nature.

Collisions with a local moonlet population present the most likely explanation and the lifetime and numbers mean that $\sim15$ mini-jet forming collisions must happen each day. Depending on assumptions about the structure of the F ring core and the nature of the collisions this implies a population of  order hundreds of $\lesssim1$km radius objects. Some of these moonlets have enough strength to survive multiple collisions but others are disrupted into groups of smaller objects on similar orbits. We suggest that they are likely formed in the F ring itself, possibly due to the action of Prometheus, and are subsequently perturbed onto colliding orbits by further interactions. Those that survive may continue to grow and go on to form the larger visible objects such as S6. Mini-jets, therefore, represent one end of a continuum of collisional features with the other end being the large jets and spiral strands.

Acknowledgments: This work was supported by the Science and Technology Facilities Council (grant number ST/F007566/1).

\begin{figure}
\begin{center}
\includegraphics[width=16.45cm]{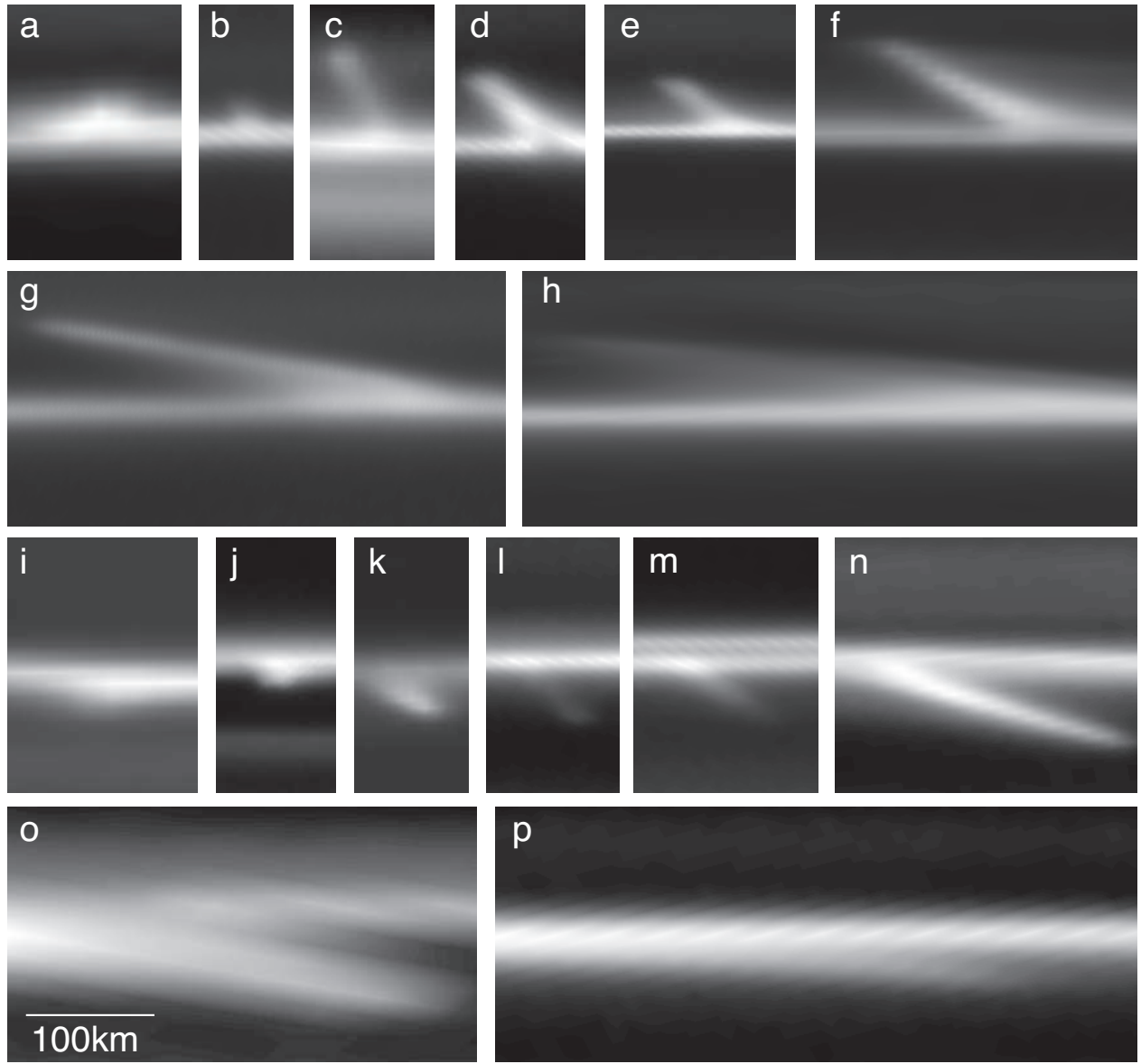}
\caption{Classic mini-jets re-projected in a radius/longitude frame to the same scale and ordered by age. Outwards mini-jets: (a) N1577813677a, (b) N1604028396, (c) N1733559846,  (d) N1613003098,  (e) N1597907705, (f) N1612005469, (g) N1623284964, (h) N1623331766c. Inwards mini-jets: (i) N1726901763a, (j) N1607629517, (k) N1605396128a, (l) N1616541581, (m) N1615488367, (n) N1610401148b,  (o) N1727800548, (p) N1734593691. Contrast has been adjusted in each case to enhance visibility.}
\label{fig1} 
\end{center}
\end{figure}

\begin{figure}
\begin{center}
\includegraphics[width=16.45cm]{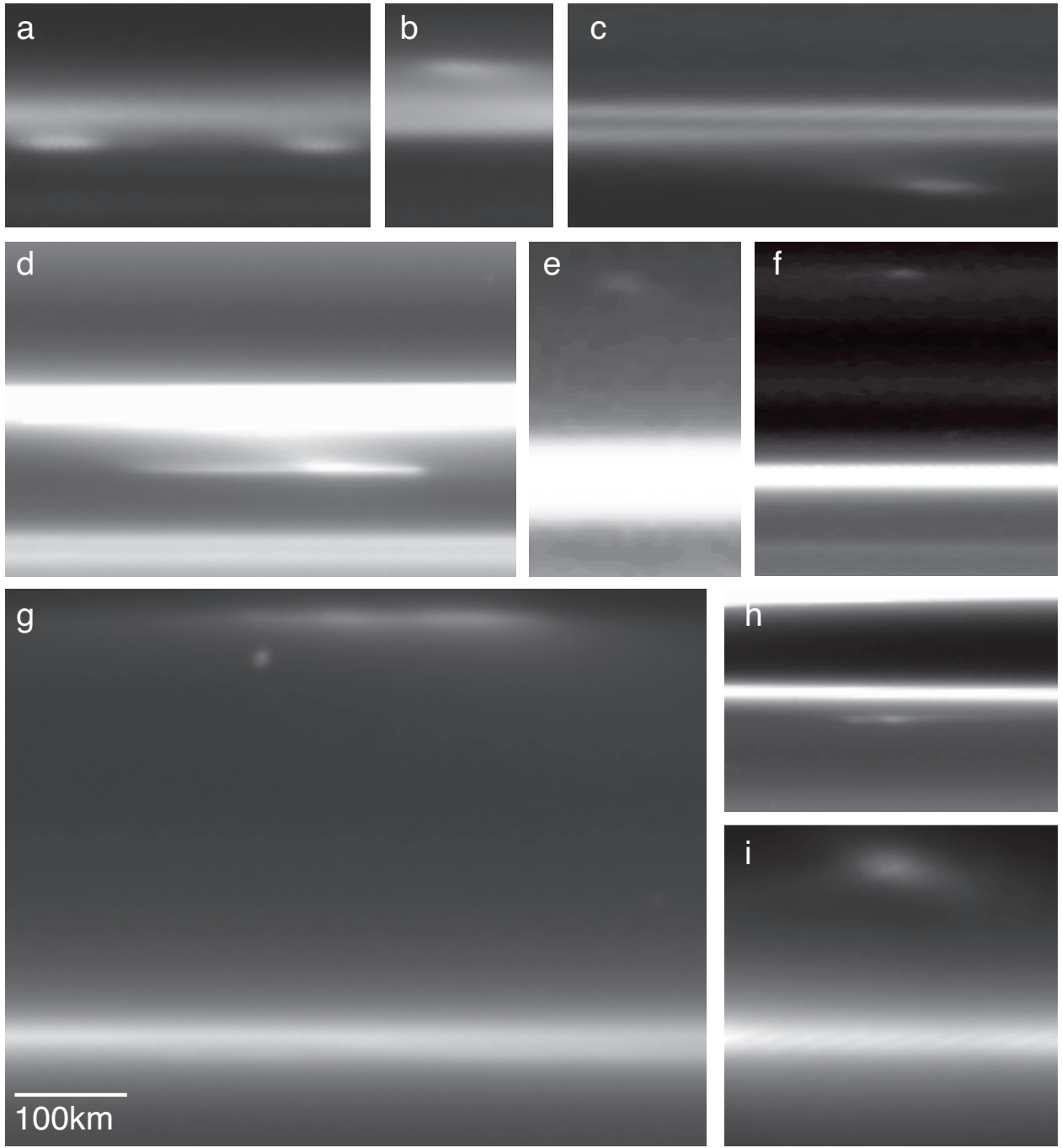}
\caption{Objects re-projected in a radius/longitude frame to the same scale. (a) multiple objects N1616523599a,b, (b) N1501710723, (c) object which could be a bright head mini-jet N1623351880, (d) N1589120162, (e) N1549820347 (f) N1610593686 which could be a clump in the outer strand, (g) large, S6-like, object N1616506283 (artefact visible below this object), (h) N1589119327b, (i) N1618601283. Contrast has been adjusted in each case to enhance visibility.}
\label{fig2}
\end{center}
\end{figure}

\begin{figure}
\begin{center}
\includegraphics[width=16.45cm]{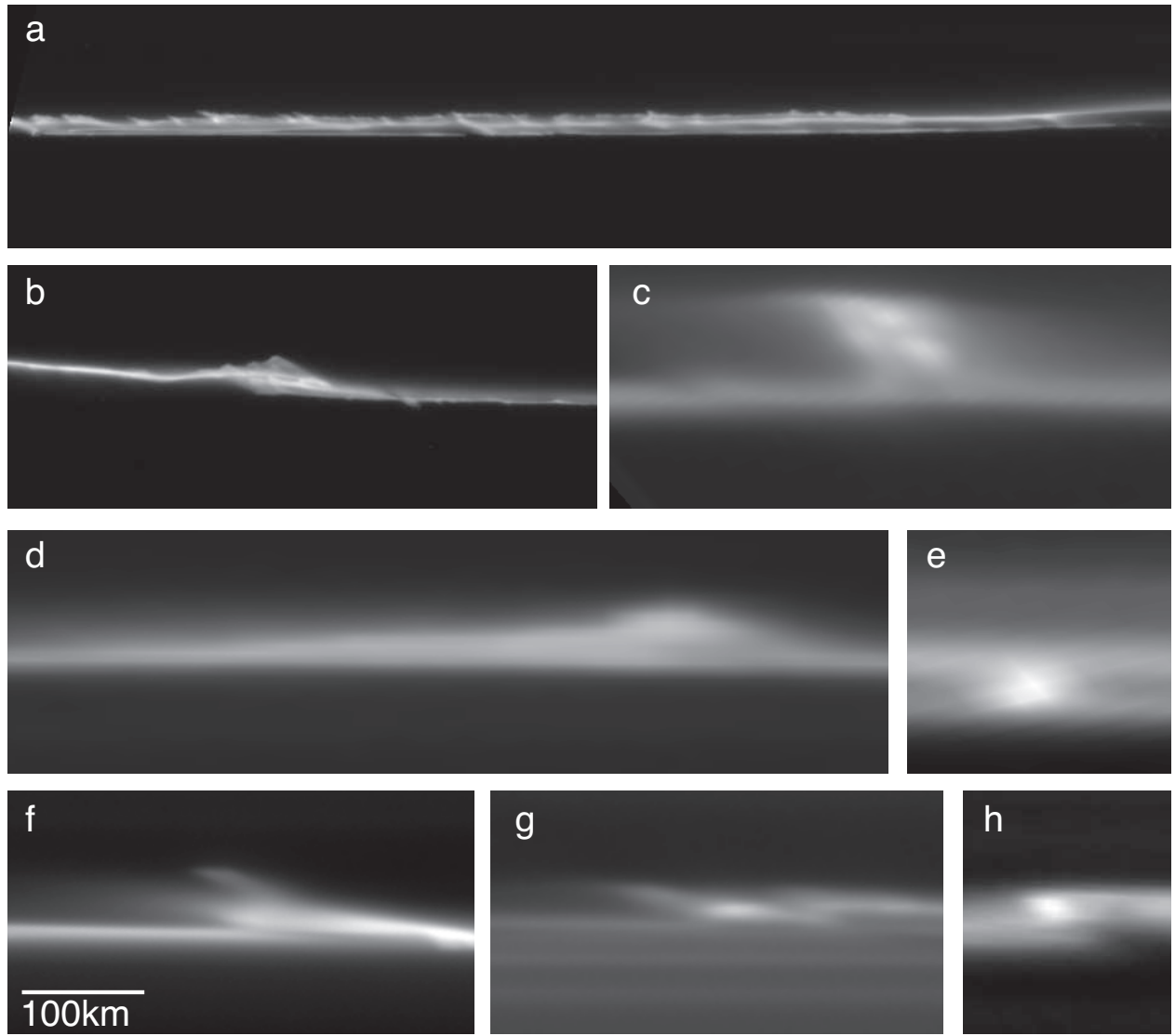}
\caption{Complex features re-projected in a radius/longitude frame to the same scale. (a) numerous small mini-jets in N1537899083 and (b) N1537898708. Complex and poorly resolved mini-jets in  (c) N1589620046 and (d) N1493639016 and possible object in (e) N1627640563. Complex mini-jets in (f) N1605531856, (g) N1623226701 and unknown feature in (h) N1601512734. Contrast has been adjusted in each case to enhance visibility. }
\label{fig3}
\end{center}
\end{figure}

\begin{figure}
\begin{center}
\includegraphics[width=16.45cm]{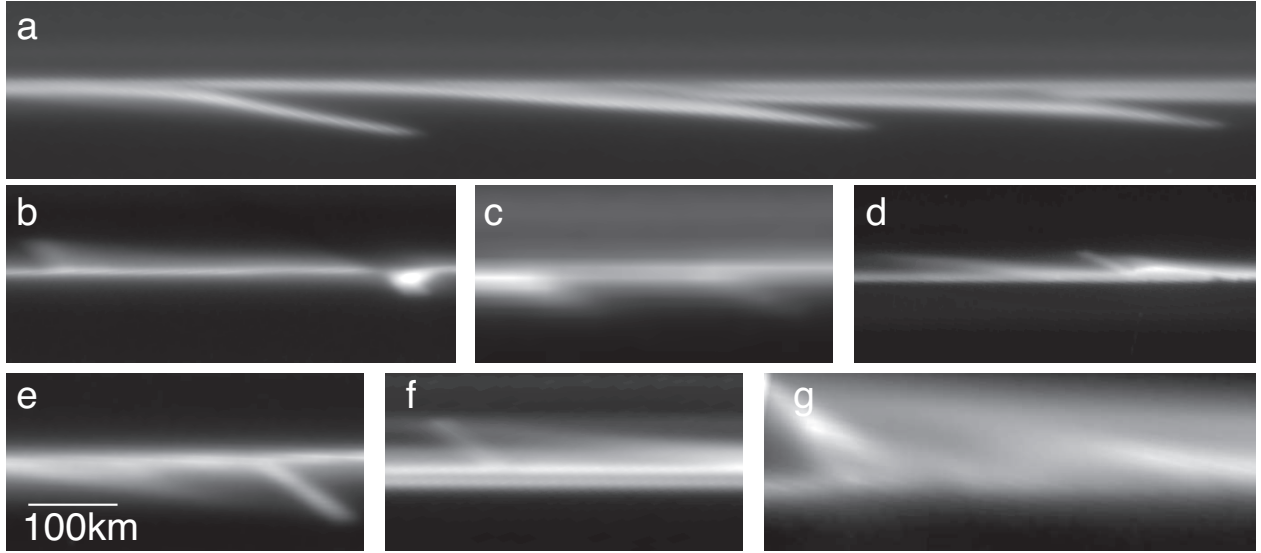}
\caption{Multiple features re-projected in a radius/longitude frame to the same scale. Mini-jets with similar ages (a) N1597902245a,b,c, (b) N1623224391a,b, (c) N1554046873a,b. Possible repeat collisions (c) N1729259467a,b,  (d) N1615511698, (e) N1727132335a,b, (g) N1733524214a,b. Contrast has been adjusted in each case to enhance visibility.}
\label{fig4}
\end{center}
\end{figure}

\begin{figure}
\begin{center}
\includegraphics[width=16.45cm]{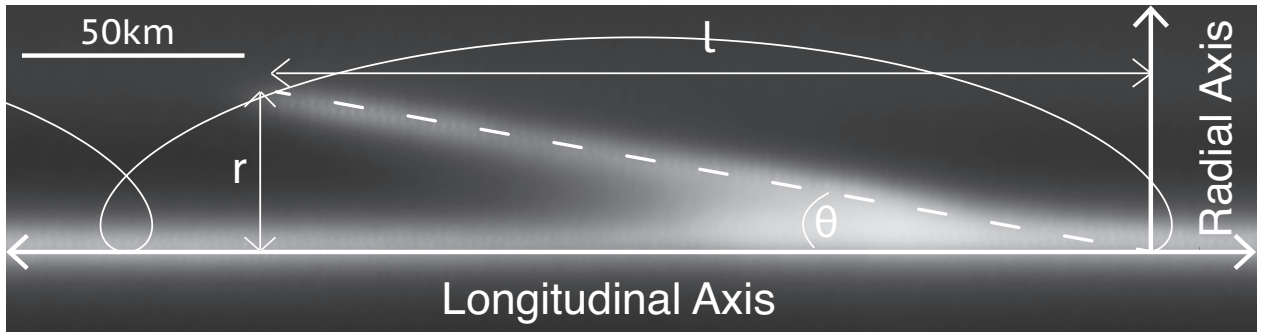}
\caption{Schematic overlaid on a typical outwards mini-jet. The origin is centred on the point of collision and the white curve is the trajectory of the tip. The mini-jet starts out pointing the `wrong way', against the direction of Keplerian shear, before quickly passing through the vertical and looping away down the ring. The colliding object has a similar orbit and will also roughly follow this path.}
\label{fig5}
\end{center}
\end{figure} 	

\begin{figure}
\begin{center}
\includegraphics[width=16.45cm]{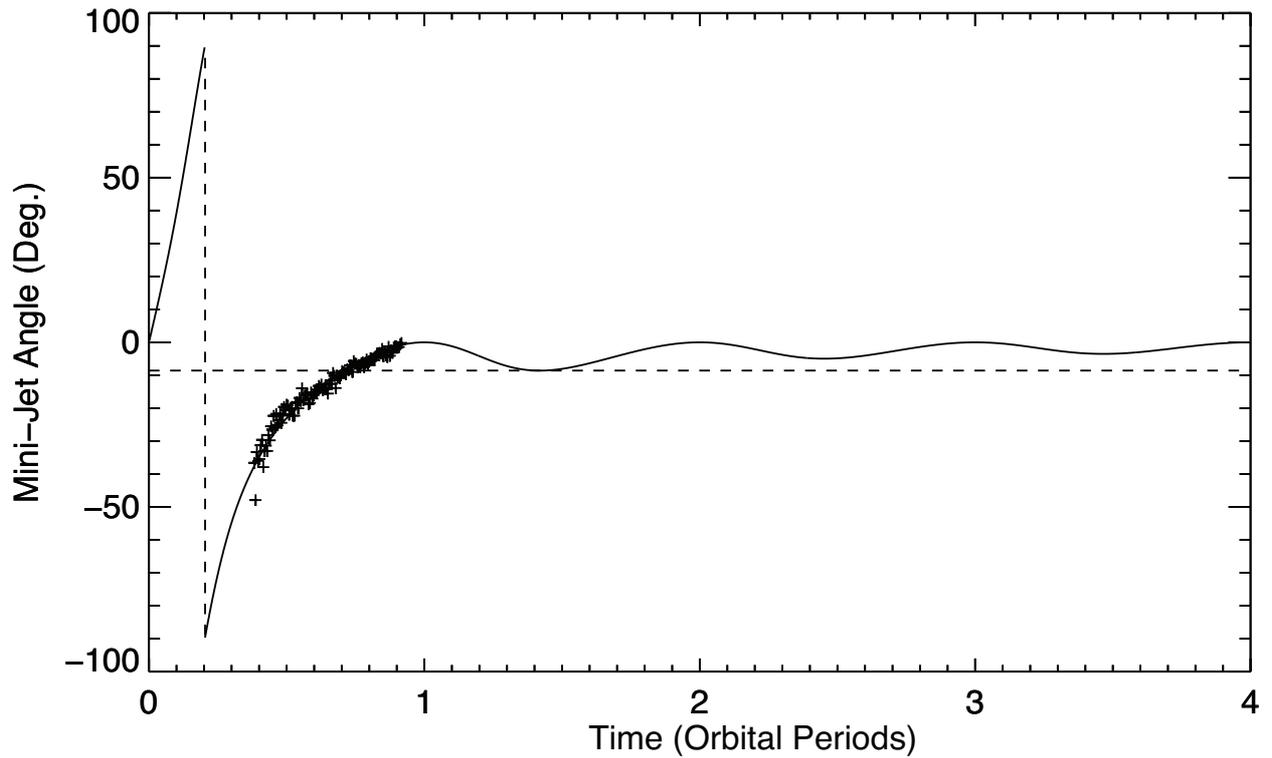}
\caption{Time evolution, in orbital periods ($14.9$ hours at the F ring), of $\theta$, the angle in degrees that a mini-jet makes with the longitudinal axis. The angle passes through the vertical (denoted by the vertical dashed line) early in the first cycle before collapsing back into the core and then oscillating in and out of the core at ever smaller angles. The horizontal dashed line at $\theta = -8.56^{\circ}$ marks the minimum of the second cycle. All mini-jets with angles less than this are on their first cycle while those with greater angles are degenerate and could be any number of cycles old. Crosses are the measured angles of the original feature over time (see \citet{attree12} and supplement B of this work for details) with errors $\sim1^{\circ}$.}
\label{fig6}
\end{center}
\end{figure}

\begin{figure}
\begin{center}
\includegraphics[width=16.45cm]{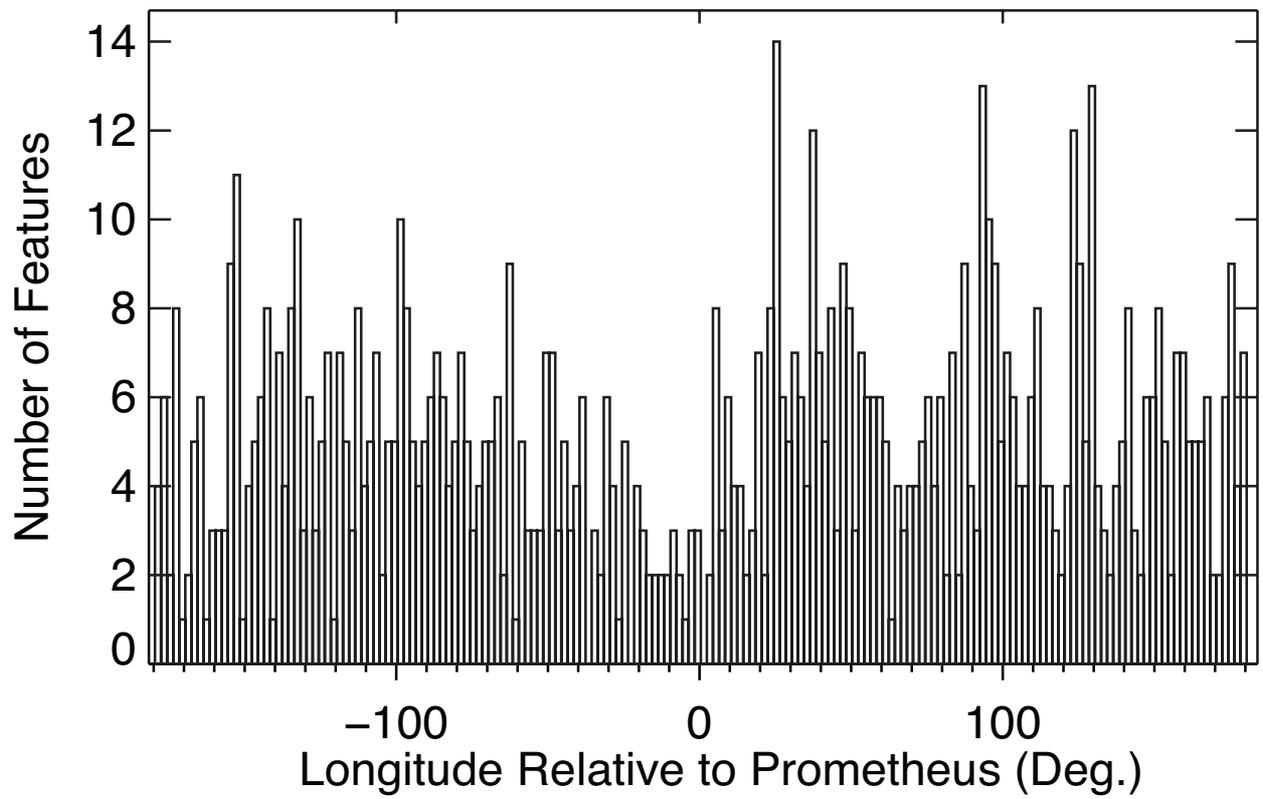}
\caption{The number of features seen in $2^{\circ}$ bins relative to Prometheus. There are no statistically significant trends.}
\label{fig7}
\end{center}
\end{figure}

\begin{figure}
\begin{center}
\includegraphics[width=16.45cm]{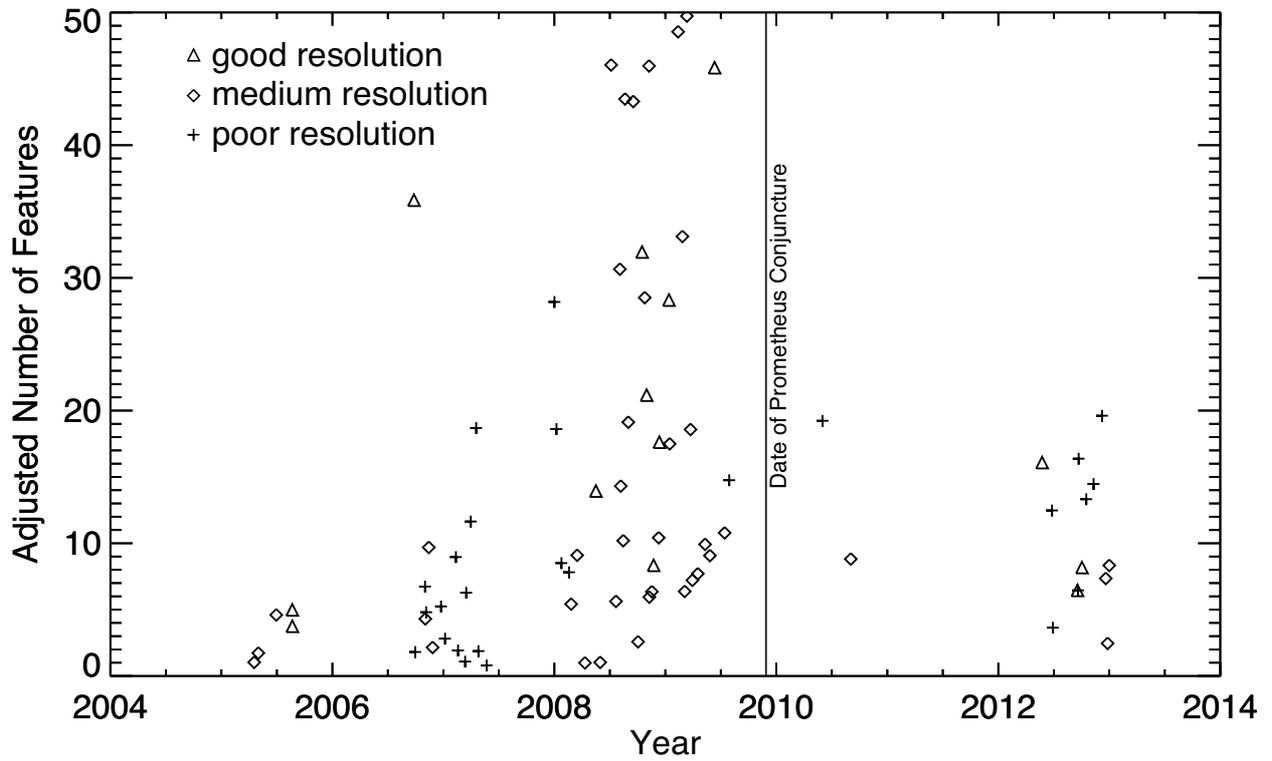}
\caption{The number of features, per observation sequence with $>50^{\circ}$ coverage, as a function of time. The number of features has been adjusted by longitude coverage as described in Section 4.1. There are no statistically significant trends. Good resolution refers to sequences imaged from a mean range of $<8\times10^{5}$km (radial resolution $\lesssim5$km per pixel), medium is between $8\times10^{5}$ and $1.6\times10^{6}$km and poor is $>1.6\times10^{6}$km (radial resolution $\gtrsim10$km per pixel).}
\label{fig8}
\end{center}
\end{figure}

\begin{figure}
\begin{center}
\includegraphics[width=16.45cm]{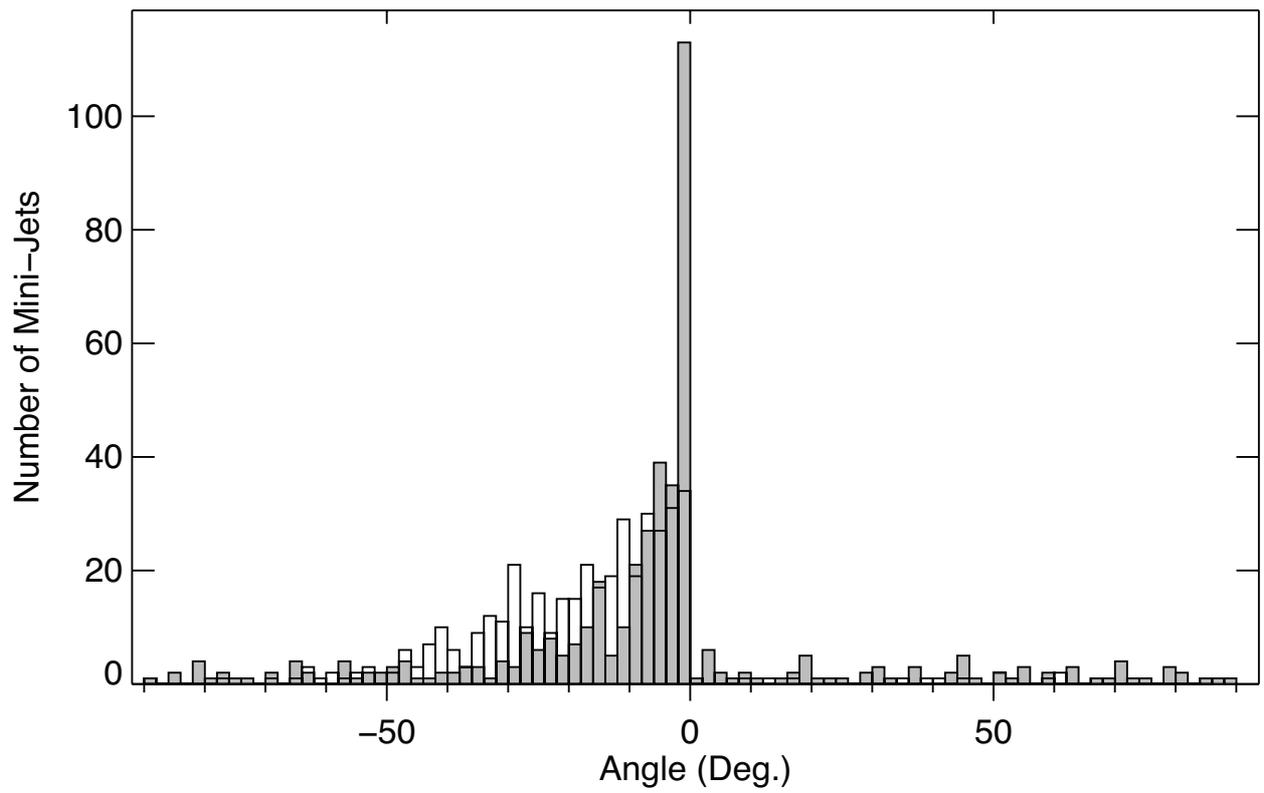}
\caption{Histogram of the measured mini-jet angles, $\theta$, in $2^{\circ}$ bins and a typical predicted distribution (shaded). See Section 4.2 for details.}
\label{fig9}
\end{center}
\end{figure}

\begin{figure}
\begin{center}
\includegraphics[width=16.45cm]{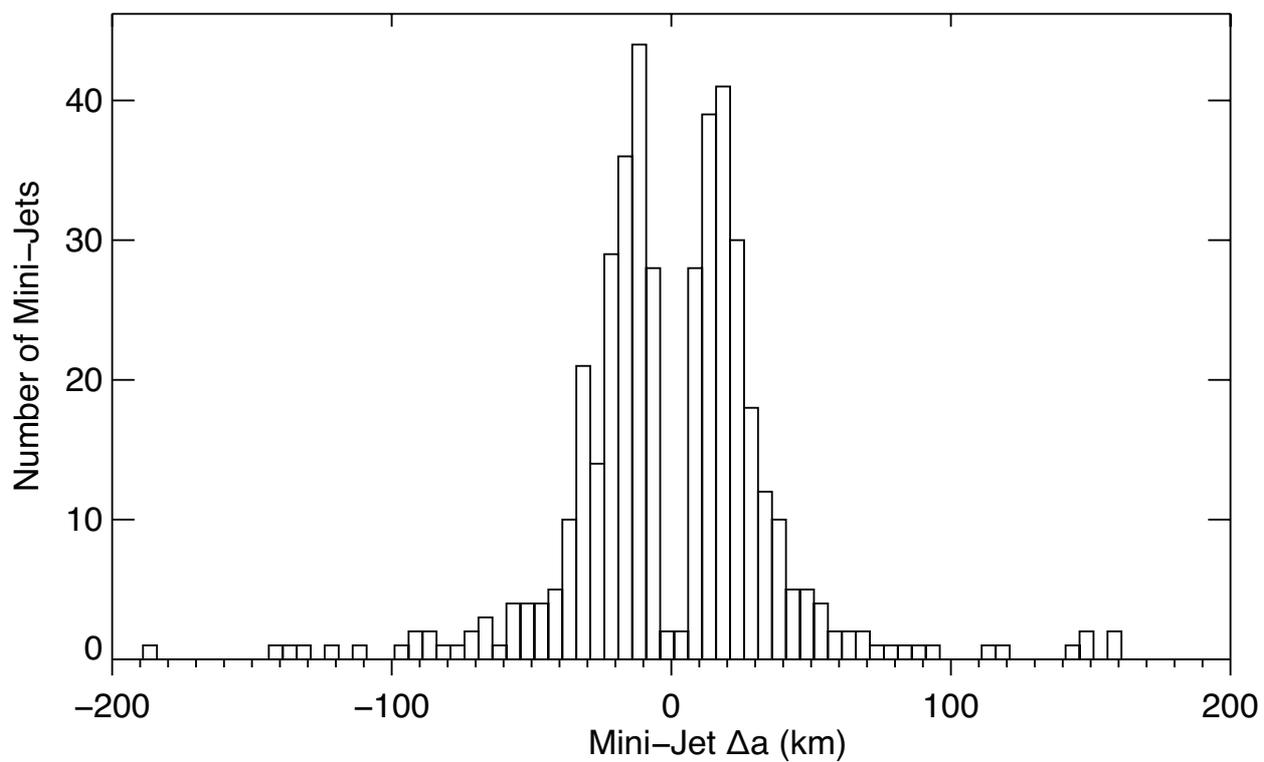}
\caption{Histogram of the calculated mini-jet $\Delta a$ (semi-major axis of the mini-jet tip relative to its base) in $5$km bins.}
\label{fig10}
\end{center}
\end{figure}

\begin{figure}
\begin{center}
\includegraphics[width=16.45cm]{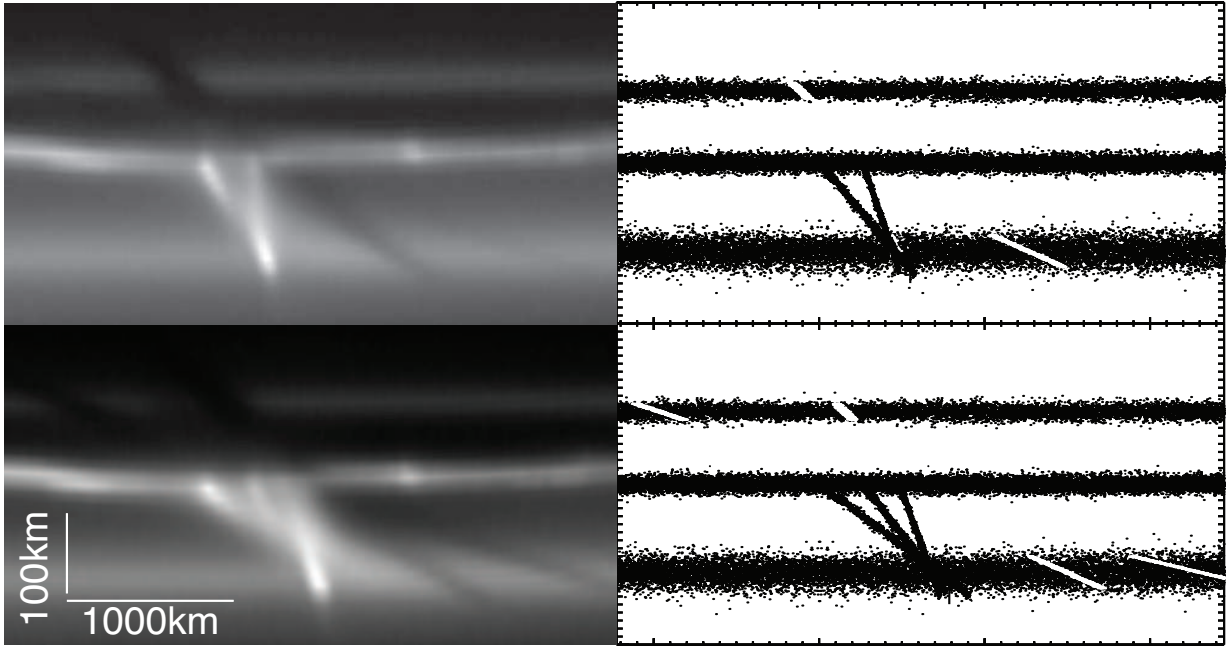}
\caption{Multiple mini-jet feature imaged twice approximately one orbital period apart. Left: re-projections of Cassini NAC image N1557026084, taken at 2007-125-02:40:53.219, and N1557080024, taken at 2007-125-17:39:52.877. See Table \ref{tab1} for further details. Right: corresponding frames from a formation model. A single object with a $\Delta a$, $\Delta e$ and $\Delta i$ relative to the F ring core, loops along from left to right, creating a new mini-jet each time it moves through the core and sweeping out a dark channel on each pass through the strands. These subsequently shear to form the observed structure.}
\label{fig11}
\end{center}
\end{figure}

\begin{figure}
\begin{center}
\includegraphics[width=16.45cm]{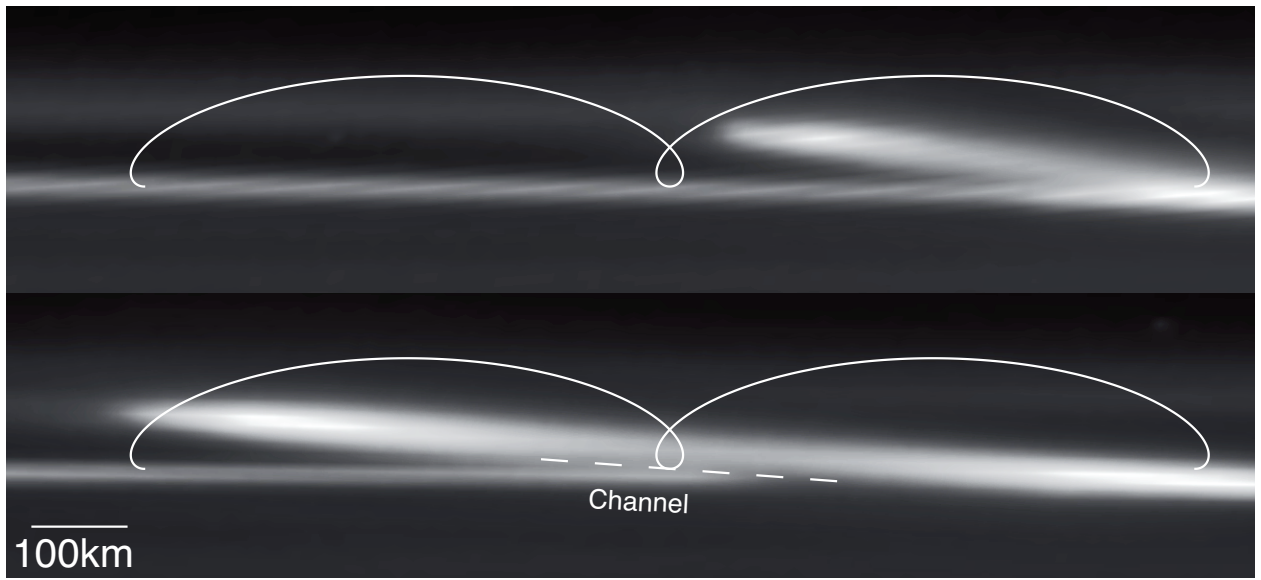}
\caption{Mini-jet imaged twice, one orbital period apart; the image numbers are top: N1739126746, taken at 2013-040-17:51:25.327 and bottom: N1739180046, taken at 2013-041-08:39:44.988. The path of the tip/object is superimposed assuming $\Delta a \approx 65$km which matches the location of the bright head in both images and the dark channel in the core in the second. This then represents an object re-colliding with a sparsely populated piece of ring and sweeping up material rather than impacting to form a second mini-jet. Compare with Fig.~\ref{fig11} above.}
\label{fig12}
\end{center}
\end{figure}

}
\end{document}